\documentclass[twocolumn]{aa}
\usepackage{graphicx}
\usepackage{txfonts}

\begin{document}

\title{The Large Scale Structure of the Galactic Magnetic Field
and High Energy Cosmic Ray Anisotropy}

\author{Jaime Alvarez-Mu\~niz 
\inst{1}
\and
Todor Stanev \inst{2}
}

\institute{Departamento de F\'\i sica de Part\'\i culas,
Universidade de Santiago de Compostela, \\ 
15782 Santiago de Compostela, A Coru\~na, SPAIN\\
\email{jaime@fpaxp1.usc.es}
\and
Bartol Research Institute, Department of Physics and Astronomy,
University of Delaware, \\
Newark, Delaware 19716, U.S.A.
}

\date{Received July 2005}

\abstract{
The magnetic field in our Galaxy is not well known and is
difficult to measure. A spiral regular field in the disk 
between the Galactic arms is
favored by observations, however it is still controversial if the field
reverses from arm to arm. The parity of the field across the 
Galactic plane is also not well established. 
In this letter we demonstrate that 
cosmic ray protons in the energy range $10^{18}$ to $10^{19}$ eV,
if accelerated near the center of the Galaxy,
can probe the large scale structure of the Galactic Magnetic Field.
In particular if the field is of even parity, 
and the spiral field reverses direction from arm to arm, i.e. 
if it is bi-symmetric (BSS), ultra high energy protons
will predominantly come from the Southern Galactic hemisphere,  
and predominantly from the Northern Galactic
 hemisphere if the field is of even parity and axi-symmetric (ASS).
 There is no sensitivity
to the BSS or ASS configurations if the field is of odd parity.  
\keywords{cosmic rays -- magnetic fields}
}

\authorrunning
\titlerunning

\maketitle

\section{\label{intro}Introduction}

The Galactic Magnetic Field (GMF) is not well known and is hard to study
(\cite{Han:2002,Vallee:2004,Wielebinski:2005}). 
 The position of the Solar system makes it
 difficult to measure its global structure and to distinguish local
small-scale features from large-scale ones. Faraday rotation measures (RM)
of pulsars in our Galaxy and of polarized extragalactic radio sources are
one of the best probes of the large scale structure of the GMF in the
Galactic disk and the halo.  From this measurements it is derived that the
GMF has two components: a regular component with strength $\sim $ few
$\mu$G, and a turbulent or random component of the same or perhaps even
larger strength (\cite{Beck:2001}). There seems to be agreement on the 
spiral structure of
the regular field in the Galactic plane between the Galactic arms,
 although not on the exact shape of the spiral field, axi-symmetric (ASS)
 or bi-symmetric (BSS) (\cite{Han:2003,Vallee:2004}). There is some 
controversy on the number of field reversals from arm to arm.
The controversy extends to the parity (even or odd) of the GMF across the
Galactic plane.  There is also disagreement on the existence of a possible halo
field.
An A0 dipole field directed towards the North Galactic Pole (NGP) was suggested
 as a halo field (\cite{Han:1997}). 

Cosmic ray propagation in the Galaxy is strongly affected by the GMF. The
gyroradius of a proton of energy $E=10^{18}$ eV in a $3~\mu$G field
is of the order of 300 pc, the typical thickness of the Galactic disk.  For
energies $E<10^{18}$ eV cosmic rays diffuse in the GMF, they get
isotropized, and hence do not reveal the sources where they were
 accelerated. Such cosmic rays are also fairly insensitive to the
 large features of the poorly-known GMF. 
 At energies above $10^{19}$ eV cosmic rays have long been thought to
 be of extragalactic origin because there are no astrophysical objects
 with high magnetic fields on the large scale needed for their 
 acceleration (\cite{Cocconi:1956}). Even if their sources were 
 inside the Galaxy,
there would exist a clear anisotropy in the arrival direction of cosmic
rays in the case of protons that is not supported by data.  Heavier nuclei
such as iron would not be ruled out as having a Galactic origin, because
 with their lower rigidity they would become isotropized in the GMF
 even in this energy range. However, their presence in the cosmic
 ray spectrum above $10^{19}$ eV is less favored by composition
measurements (\cite{FlysEye:2005}). An extremely interesting energy range is 
that from $10^{18}$
to $10^{19}$ eV. In this energy bin cosmic ray propagation through the GMF
is thought to change from diffusive to ballistic, cosmic ray composition is
thought to change from heavy to light, and cosmic ray origin is thought to
change from Galactic to extragalactic (\cite{Nagano:2000}).   

The center of our Galaxy, where there is some evidence for the existence of
a very massive black hole, provides a natural candidate for acceleration of
cosmic rays to very high energies. 
 The high energy astrophysical activity at the Galactic center is 
 supported by the recent 
observation by the HESS telescope of a TeV gamma-ray source near the
location of Saggitarius $A^*$ (\cite{HESS:2004}).  

 In this work we demonstrate that protons in the energy range
 from $10^{18}$
 to $10^{19}$ eV, if accelerated at galactocentric distances
typically smaller than the radius of the Solar system orbit around the 
Galactic center,
are sensitive to the large scale structure of the GMF. In
particular, if the GMF is of even parity, i.e. does not change sign across
the Galactic plane, their distribution in arrival direction reveals the
axi-symmetric (ASS) or bi-symmetric (BSS) configuration of the spiral field. In
the first case we will show that protons are observed to arrive
predominantly from the Northern Galactic hemisphere, while in the second
they are seen to come predominantly from the Southern Galactic hemisphere. 
However, as we demonstrate below, there is no sensitivity to the GMF
configuration if the field changes sign across the Galactic plane,
i.e. if it is of odd parity.   

This letter is structured as follows: In Section \ref{GMF} we briefly review
the current knowledge on the GMF, and describe the GMF configurations used
in this work. In Section \ref{technique} we briefly describe
how we performed our calculations.
In section \ref{results} we demonstrate the sensitivity of 
cosmic ray propagation to the large scale features of the GMF. 

\section{\label{GMF} The Galactic Magnetic Field}

We briefly summarize here the current knowledge
on the GMF. More details can be found in 
(\cite{Han:2003,Vallee:2004,Wielebinski:2005}).
At low Galactic latitudes 
Faraday RMs of pulsars inside the Galaxy 
reveal that the disk field direction is coherent over a linear scale
of at least a few kpc between  
the Galactic arms. It is not yet clear from an experimental 
point of view if the field 
reverses direction from arm to arm (\cite{Han:2003,Vallee:2004}). 
In particular, moving
towards the Galactic Center (GC) the field direction -- clockwise-going or
anticlockwise-going as seen from the North Galactic Pole -- between
the Perseus and Perseus + I arm at galactocentric distance 
$r_{\vert\vert}\sim 12$ kpc is still controversial.
There seems to be agreement on the existence of a clockwise-going field 
between the Perseus and Carina-Sagittarius arm
at $r_{\vert\vert} \sim 8$ kpc,  
close to the position of the Solar System, 
and with a strength of $B\approx 2 \dots 4~\mu$G.
The field reverses direction in the next arm
towards the Galactic Center (GC) between the Carina-Sagittarius
and the Crux-Scutum arms at $r_{\vert\vert}\sim 6.5$ kpc. 
Moving closer to the GC, 
the field is clockwise-going between
the Crux-Scutum and Norma arms at $r_{\vert\vert}\sim 5$ kpc. Finally
near the Norma arm at $r_{\vert\vert}\sim 4$ kpc measurements are again 
contradictory.

\begin{figure*}
\centering
\includegraphics[width=8.5cm]{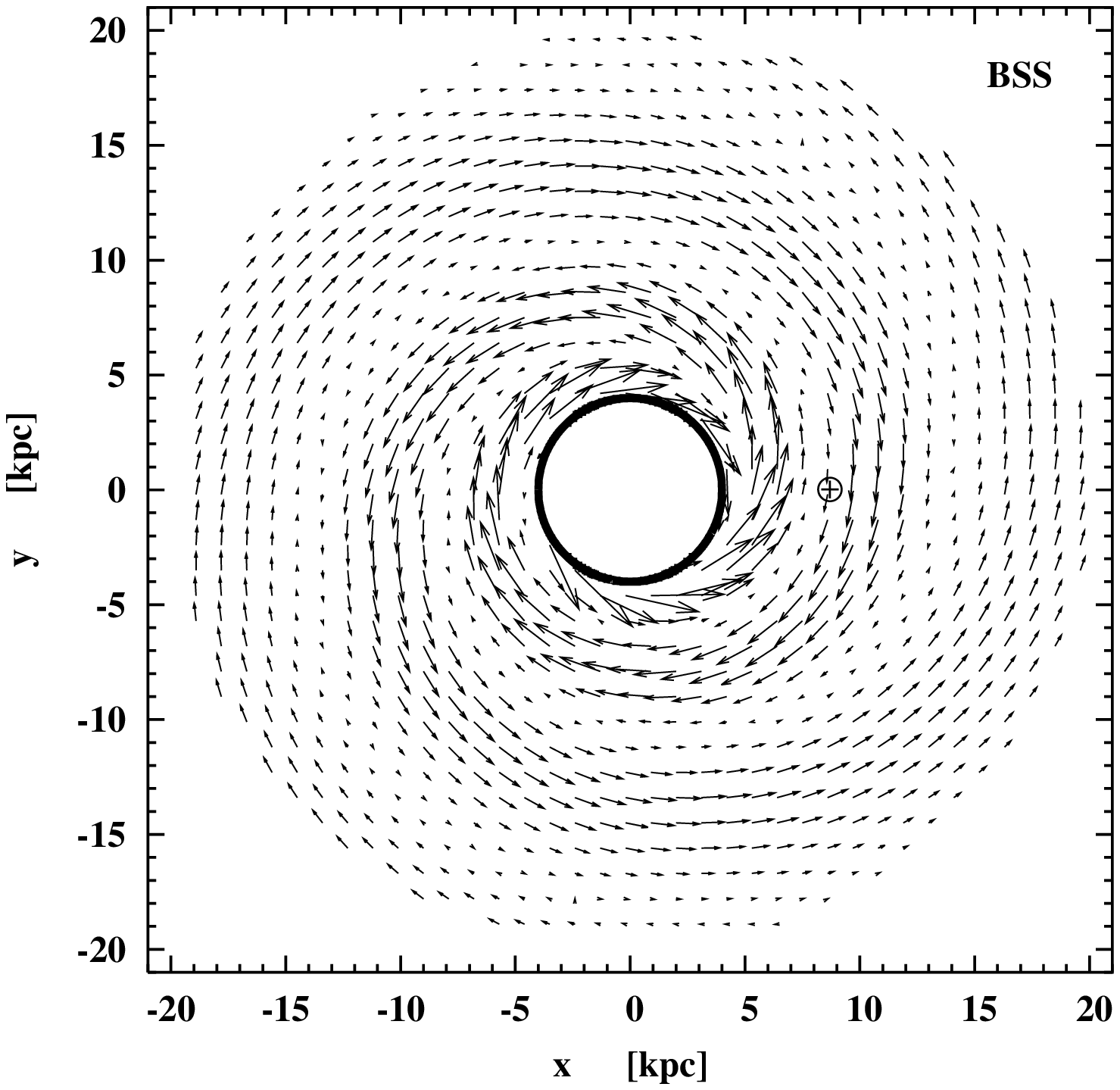}
\includegraphics[width=8.5cm]{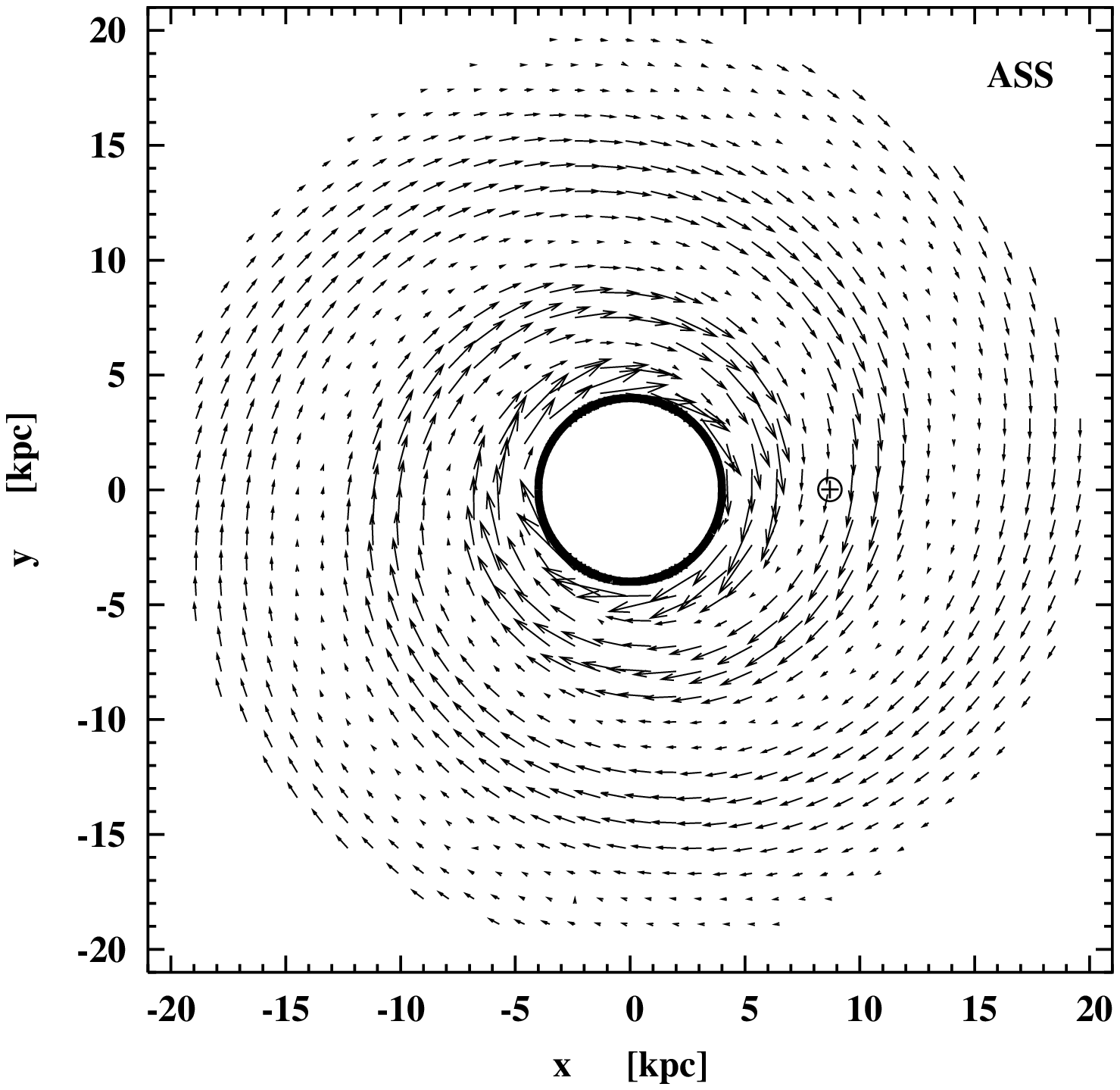}
\caption{Spiral component of the regular GMF in the Galactic plane.
The vectors indicate the field direction and their length is
proportional to its magnitude. Left panel: Bi-symmetric spiral field.
Right panel: Axi-symmetric spiral field without reversals.
The position of the Solar
system is indicated with an open circle with a cross inside.
The solid thick circle
is the ring in the Galactic plane of radius $r=4$ kpc
around the Galactic center where the sources are located.
The magnetic field lines inside this circle are not plotted for
clarity purposes.
}
\label{fig:spiral-field}
\end{figure*}

All this evidence gives support for a two-arm logarithmic
spiral regular field in the disk. It is  
not well established whether the disk field is better described by 
a bi-symmetric spiral (BSS) configuration which allows for multiple 
field reversals, or an axi-symmetric spiral (ASS) configuration,
with the BSS structure slightly favored. These two field configurations
are plotted in Fig.\ref{fig:spiral-field}.

Also it has recently been confirmed that the GC contains a 
highly regular polar field consistent with the presence of 
a dipole field in the Galaxy and compatible with being
generated by an A0 dynamo (\cite{Han:2003}). The presence of the dipole field 
could explain  
the vertical component of the GMF of strength $B_z \approx 0.3 \pm 0.1~\mu$G 
observed in the vicinity of the Sun. 

At high Galactic latitudes the antisymmetric
distribution of Faraday RMs is indicative of the
odd parity of the field although an even parity is not 
excluded. At large distances from the disk $> 1$ kpc or so,
the measurements are again very difficult because 
of contamination by the GMF in the disk.

Given this information, in this paper we do not make 
strong assumptions about the large scale structure of the
spiral GMF and assume all possible
 combinations, namely ASS without field reversals or BSS,
 and even or odd parity. 
The local regular magnetic field in the vicinity of the Solar System
is assumed to be
$\sim 1.5~\mu{\rm G}$ in the direction $l=90^{\rm o}+p$ where the
pitch angle is $p=-10^{\rm o}$ (\cite{Han:1994}). Some measurements
discuss larger total field strengths of up to 6 $\mu$G (\cite{Beck:2001}),
therefore our assumptions should be considered as rather conservative.
The field decreases with Galactocentric distance as $1/r_{\vert\vert}$
and it is zero for $r_{\vert\vert}>20$ kpc.
In the region around the Galactic center ($r_{\vert\vert} < 4$ kpc)
the field is highly uncertain, and 
we assume it is constant and equal to its
value at $r_{\vert\vert}=4$ kpc.
Following (\cite{Stanev:1996})
the spiral field strengths above and below the Galactic plane are taken to
decrease exponentially with two scale heights.
In this work we also assume a halo field corresponding to an
A0 dipole as suggested by (\cite{Han:2002a}).
The dipole field is toroidal and its strength
decreases with Galactocentric distance as $1/r^3$.
The dipole field is very strong in the central region
of the Galaxy, but is only 0.3 $\mu$G in the vicinity
of the Solar system, directed towards the North Galactic Pole.
The equations describing the functional form of the field strength
for both the spiral and the dipole fields have been published elsewhere 
(\cite{Stanev:1996,Alvarez-Muniz et al:2002,Prouza:2003}).

We also assume a significant turbulent component of the GMF, $B_{\rm ran}$.
Its strength is comparable or possibly larger than the regular field 
(\cite{Beck:2001}).
 To simulate it we add to the spiral and dipole components
a random field with a strength of 50\% of the local regular field
strength with coherence length of 100 pc. We have also 
performed several runs with a turbulent field twice the local 
regular field (i.e. four times the standard random field).
  The possible geometrical offset of the random
 and regular fields and the possible time dependence of $B_{\rm ran}$ are
 neglected.

\section{\label{technique} Calculation technique}

We sample protons with energies greater than $E=10^{18}$ eV
from a $dN/dE \propto E^{-2.7}$
energy spectrum. We inject 100 protons per source isotropically from
sources distributed homogeneously in a ring of radius $r_{\vert\vert}=4$ kpc
around the Galactic center in the Galactic plane. We 
forward (not backward) propagate them from the sources 
in different models of the GMF 
by numerically integrating the equations of motion in a
magnetic field.
 There is no energy loss on propagation.
 We stop the propagation and sample a new
proton energy when the proton trajectory intersects our detector 
-- a 1 kpc radius sphere around the
Solar system position -- when it reaches
Galactocentric distances $r_{\vert\vert}>20$ kpc, or when 
it travels a
total pathlength larger than 4 Mpc. The total length of trajectories
 reaching Earth is always much smaller than this limit.
 If a proton hits the detector, we keep the proton arrival
 direction in Galactic coordinates, as well as the position
 of the source from which it was injected.
 Our results can be easily re-scaled in rigidity for heavier nuclei.
 We will show in the next section that our conclusions do not
 change qualitatively if we reduce the radius of the detector
 around the Solar System, although
the detection efficiency -- the ratio of detected to
injected protons -- decreases as the area of the spherical
detector. For this reason we prefer to use a relatively large detector. 

 Given the still controversial experimental measurements of the GMF,
 we study the sensitivity of cosmic ray propagation to an ASS and a
 BSS spiral field, that can be either of even or odd parity across
 the Galactic plane. In order to better understand the effect of
 each of the GMF components -- spiral, random and dipole -- we simulate
 the propagation of cosmic rays artificially switching on each component,
 starting with the spiral field, then adding the random component
 and finally switching all of them on. A total number of 50,000 protons
 are collected for each configuration of the GMF we have explored.

 Numerical Monte Carlo procedures such as the one described above
 do not usually represent well the power spectrum of the random
 magnetic field and thus can not be used to predict the energy
 dependence of the diffusion coefficient. In this study we are 
 concentrating on the propagation of protons in the transition
 regime  between diffusive and ballistic propagation, where the 
 random field properties are less important.

\section{\label{results} Results}

Tables 
\ref{tab:NS_BSS_even}, \ref{tab:NS_ASS},
and \ref{tab:NS_BSS_even_double} 
summarize the main results of our work. In all of them we give the 
fraction of cosmic rays in different energy
bins coming from the Southern Galactic hemisphere (SGH), i.e.
 that arrive at the spherical detector from Galactic latitude $b<0$.
 The numbers in parenthesis are the one sigma Poisson 
 limits on the fraction of events, i.e. the probability that the 
fraction of events is outside the corresponding interval is $1/e \sim 0.37$.  
 Typically the uncertainty increases with energy as less events
 are sampled at high energy due to the steep injection spectrum. 

\begin{table*}
\caption{Fraction of protons in different energy bins
that arrive from the Southern Galactic hemisphere (SGH)
at our spherical detector around the Solar system
after propagating through
the BSS even parity GMF model. We inject 100 protons
per source with the sources homogeneously distributed in the
Galactic plane in a
ring of radius $r_{\vert\vert}=4$ kpc around the Galactic center.
The fraction of events when switching on different components of
the GMF is shown.
Second column: Spiral field only. Third column: Spiral and
random field. Fourth column: Spiral and random and dipole field.
 The numbers in parenthesis are the one sigma Poisson
 limits on the fraction of events (see text).}
\label{tab:NS_BSS_even}
\renewcommand{\arraystretch}{1.5}
\centering
\begin{tabular}{  c   c   c   c  }
\hline
 & \multicolumn{3}{c  }{~Even parity~} \\ \cline{2-4}

~$\log_{10}(E/{\rm eV})$ bin~ & ~BSS only~ & ~BSS + B$_{\rm ran}$~ & BSS +
A0 +
B$_{\rm ran}$ \\ \hline

  18.0 - 18.1  & ~$99.0 ~(100.0,~95.4)~ $~  & ~$95.3 ~(95.5,~95.1)~$~  &
~$70.0 ~(70.2,~69.8)~$~ \\
  18.1 - 18.2  & ~$99.1  ~(100.0,~95.5)~$~  & ~$96.3 ~(96.6,~96.0)~$~  &
~$84.7 ~(85.0,~84.4)~$~ \\
  18.2 - 18.3  & ~$100.0  ~(100.0,~99.8)~$~  & ~$96.3 ~(96.7,~95.9)~$~  &
~$94.9 ~(95.4,~94.5)~$~ \\
  18.3 - 18.4  & ~$100.0 ~(100.0,~99.7)~$~  & ~$97.6 ~(98.6,~96.6)~$~  &
~$98.6 ~(99.2,~98.0)~$~ \\
  18.4 - 18.5  & ~$100.0 ~(100.0,~99.5)~$~  & ~$98.2 ~(100.0,~96.2)~$~  &
~$99.5 ~(100.0,~98.9)~$~ \\
  18.5 - 18.6  & ~$100.0 ~(100.0,~96.5)~$~ & ~$96.7 ~(100.0,~92.9)~$~  &
~$99.9 ~(100.0,~99.1)~$~ \\
  18.6 - 18.7  & ~$100.0 ~(100.0,~94.8)~$~ & ~$95.6 ~(100.0,~89.9)~$~ &
~$100.0 ~(100.0,~98.8)~$~ \\
  18.7 - 18.8  & ~$100.0 ~(100.0,~94.4)~$~  & ~$100.0 ~(100.0,~94.0)~$~ &
~$100.0 ~(100.0,~98.4)~$~ \\
  18.8 - 18.9  & ~$100.0 ~(100.0,~92.8)~$~ & ~$100.0 ~(100.0,~93.7)~$~ &
~$100.0 ~(100.0,~98.0)~$~ \\
  18.9 - 19.0  & ~$100.0 ~(100.0,~88.9)~$~  & ~$100.0 ~(100.0,~89.7)~$~  &
~$98.2 ~(100.0,~95.9)~$~ \\ 
\hline
\end{tabular}
\end{table*}

\begin{table*}
\caption{Same as in table \ref{tab:NS_BSS_even} for the
ASS even parity GMF model (top half of the table) and for
the ASS odd parity GMF model (bottom half of the table).}
\label{tab:NS_ASS}
\renewcommand{\arraystretch}{1.5}
\centering
\begin{tabular}{  c   c   c   c  }
\hline

 & \multicolumn{3}{c  }{~Even parity~} \\ \cline{2-4}

~$\log_{10}(E/{\rm eV})$ bin~ & ~ASS only~ & ~ASS + B$_{\rm ran}$~ & ~ASS + A0 +
B$_{\rm ran}$~  \\ \hline

  18.0 - 18.1  & ~$57.5 ~(58.7,~56.4)~$~  & ~$87.3 ~(87.4,~87.0)~$~  &
~$72.6 ~(72.7,~72.4)~$~ \\
  18.1 - 18.2  & ~$36.5 ~(37.4,~35.6)~$~  & ~$84.2 ~(84.5,~83.9)~$~  &
~$56.9 ~(57.2,~56.6)~$~ \\
  18.2 - 18.3  & ~$97.8 ~(98.0,~97.5)~$~  & ~$73.3 ~(73.8,~72.9)~$~  &
~$29.1 ~(29.4,~28.9)~$~ \\
  18.3 - 18.4  & ~$97.8 ~(98.0,~97.6)~$~  & ~$49.0 ~(49.5,~48.5)~$~  &
~$15.1 ~(15.3,~14.8)~$~ \\
  18.4 - 18.5  & ~$87.5 ~(88.0,~87.0)~$~  & ~$31.8 ~(32.5,~31.1)~$~  &
~$4.8 ~(5.1,~4.5)~$~ \\
  18.5 - 18.6  & ~$25.4 ~(26.4,~24.6)~$~  & ~$26.4 ~(27.6,~25.4)~$~  &
~$0.9 ~(1.4,~0.6)~$~ \\
  18.6 - 18.7  & ~$3.2 ~(4.0,~2.6)~$~   & ~$2.5 ~(3.4,~1.9)~$~   &  ~$0.0
~(2.3,~0.0)~$~ \\
  18.7 - 18.8  & ~$0.0 ~(0.8,~0.0)~$~   & ~$0.0 ~(0.9,~0.0)~$~   &  ~$0.0
~(0.2,~0.0)~$~ \\
  18.8 - 18.9  & ~$0.0 ~(2.8,~0.0)~$~   & ~$0.0 ~(2.5,~0.0)~$~   &  ~$0.0
~(0.1,~0.0)~$~ \\
  18.9 - 19.0  & ~$0.0 ~(3.7,~0.0)~$~   & ~$0.0 ~(4.2,~0.0)~$~   &  ~$0.0
~(0.2,~0.0)~$~ \\ \hline

 & \multicolumn{3}{c  }{~Odd parity~} \\ \cline{2-4}

~$\log_{10}(E/{\rm eV})$ bin~ & ~ASS only~ & ~ASS + B$_{\rm ran}$~ & ~ASS + A0 +
B$_{
\rm ran}$~  \\ \hline

  18.0 - 18.1  & ~$50.6 ~(50.8,~50.5)~$~  & ~$50.4 ~(50.6,~50.2)~$~ &
~$49.9 ~(50.1,~49.8)~$~ \\
  18.1 - 18.2  & ~$52.1 ~(52.3,~51.8)~$~  & ~$49.9 ~(50.1,~49.7)~$~ &
~$49.1 ~(49.3,~49.0)~$~ \\
  18.2 - 18.3  & ~$45.8 ~(46.1,~45.5)~$~  & ~$50.0 ~(50.3,~49.7)~$~ &
~$50.3 ~(50.6,~50.0)~$~ \\
  18.3 - 18.4  & ~$51.8 ~(52.2,~51.4)~$~  & ~$49.5 ~(49.9,~49.1)~$~ &
~$49.8 ~(50.2,~49.4)~$~ \\
  18.4 - 18.5  & ~$51.8 ~(52.2,~51.3)~$~  & ~$53.2 ~(53.7,~52.7)~$~ &
~$48.0 ~(48.5,~47.5)~$~ \\
  18.5 - 18.6  & ~$47.7 ~(48.5,~47.0)~$~  & ~$48.6 ~(49.4,~47.9)~$~ &
~$52.8 ~(53.5,~52.1)~$~ \\
  18.6 - 18.7  & ~$49.5 ~(50.3,~48.7)~$~  & ~$48.9 ~(49.8,~48.1)~$~ &
~$49.9 ~(51.0,~48.9)~$~ \\
  18.7 - 18.8  & ~$55.3 ~(56.5,~54.2)~$~  & ~$54.0 ~(55.3,~52.9)~$~ &
~$45.1 ~(46.5,~43.9)~$~ \\
  18.8 - 18.9  & ~$51.1 ~(52.8,~49.6)~$~  & ~$50.8 ~(52.6,~49.2)~$~ &
~$55.4 ~(57.1,~53.7)~$~ \\
  18.9 - 19.0  & ~$42.3 ~(44.6,~40.3)~$~  & ~$43.4 ~(45.7,~41.4)~$~ &
~$53.2 ~(56.1,~50.8)~$~ \\ \hline

\end{tabular}
\end{table*}


Inspection of the tables leads to several important conclusions:

\begin{enumerate}

\item
If the GMF is of even
parity (Table \ref{tab:NS_BSS_even} and 
top half of Table \ref{tab:NS_ASS}), there is 
a very strong North-South (NS) anisotropy in the arrival direction of protons. 
In particular, 
for energies above $\sim 3 \times 10^{18}$ eV, more than $90\%$
of the protons that are detected come from the SGH in the BSS model, and 
from the Northern Galactic hemisphere (NGH) in the ASS model. 
This is a clear tendency that does not depend very much on the 
strength of the random or dipole components of the field as 
can be seen in Table \ref{tab:NS_BSS_even_double}. In this 
table we give the 
fraction of events coming from the SGH for the BSS even parity 
model, but using a random field twice the value of the regular field, 
i.e. four times the standard random field,
and also doubling the strength of the dipole field with respect to its  
nominal value of $0.3~\mu$G at the position of the Solar system.  

This result is very stable, and mostly dependent on the
 model of the regular field. The insufficient representation
 of the turbulence of the random field thus does not affect
 the conclusions of this study.

\item
If the field is of odd parity (bottom half of Table \ref{tab:NS_ASS}) 
there is no sensitivity to the ASS or BSS character of the 
spiral field. About half of the protons in all energy bins
come from the SGH in both the ASS (bottom half of Table \ref{tab:NS_ASS})
and BSS (values not given) odd parity configurations. The non-observation 
of a large NS anisotropy in the highest energy bins for protons
having Galactic longitude in a fairly wide angular bin 
around $l=0^\circ$ could be favoring an odd parity GMF.

\item
The NS anisotropy seen in Tables \ref{tab:NS_BSS_even}, 
\ref{tab:NS_ASS} and \ref{tab:NS_BSS_even_double} is significantly
reduced at energies below $\sim 3 \times 10^{18}$ eV, especially
in the ASS model. Clearly in the lower energy bins, proton propagation
is more affected by the random and dipole components of the GMF. 
The dipole field seems to be more important in reducing the NS
anisotropy at low energies, possibly due to its large strength
near the Galactic center (\cite{Yoshiguchi:2004}). 

\end{enumerate}

\begin{table*}
\caption{
Same as in table \ref{tab:NS_BSS_even} but this time we show
the fraction of protons coming from the SGH, with all GMF components
switched on (second column),
with $B_{\rm ran}$ twice the regular field, i.e. four times the
standard random field used in the second column (third column), and
with the dipole field twice its nominal value of $0.3~\mu$G at
the Solar system position (fourth column).}
\label{tab:NS_BSS_even_double}
\renewcommand{\arraystretch}{1.5}
\centering
\begin{tabular}{  c   c   c   c  }
\hline

 & \multicolumn{3}{c  }{~Even parity~} \\ \cline{2-4}

~$\log_{10}(E/{\rm eV})$ bin~ & BSS + A0 + B$_{\rm ran}$ & BSS + A0 + 4
B$_{\rm ran}$ & BSS + 2 A0 + B$_{\rm ran}$  \\ \hline

  18.0 - 18.1  &  ~$70.0 ~(70.2,~69.8)~$~ & ~$64.9 ~(65.1,~64.8)~$~ &
~$43.6 ~(43.7,~43.4)~$~ \\
  18.1 - 18.2  &  ~$84.7 ~(85.0,~84.4)~$~ & ~$73.0 ~(73.3,~72.7)~$~ &
~$41.0 ~(41.2,~40.8)~$~ \\
  18.2 - 18.3  &  ~$94.9 ~(95.4,~94.5)~$~ & ~$80.2 ~(80.5,~79.7)~$~ &
~$40.7 ~(40.9,~40.4)~$~ \\
  18.3 - 18.4  &  ~$98.6 ~(99.2,~98.0)~$~ & ~$86.4 ~(87.0,~85.8)~$~ &
~$44.0 ~(44.4,~43.7)~$~ \\
  18.4 - 18.5  &  ~$99.5 ~(100.0,~98.9)~$~ & ~$89.5 ~(90.2,~88.8)~$~ &
~$52.1 ~(52.6,~51.5)~$~ \\
  18.5 - 18.6  &  ~$99.9 ~(100.0,~99.1)~$ ~ & ~$93.7 ~(94.7,~92.7)~$~ &
~$57.0 ~(57.7,~56.3)~$~ \\
  18.6 - 18.7  &  ~$100.0 ~(100.0,~98.8)~$~ & ~$94.1 ~(95.6,~92.8)~$~ &
~$77.2 ~(78.7,~75.8)~$~ \\
  18.7 - 18.8  &  ~$100.0 ~(100.0,~98.4)~$~ & ~$93.7 ~(95.5,~91.9)~$~ &
~$92.9 ~(94.7,~91.1)~$~ \\
  18.8 - 18.9  &  ~$100.0 ~(100.0,~98.0)~$~ & ~$92.3 ~(94.8,~90.1)~$~ &
~$89.3 ~(91.1,~87.6)~$~ \\
  18.9 - 19.0  &  ~$98.2 ~(100.0,~95.9)~$~ & ~$90.2 ~(93.1,~87.5)~$~ &
~$92.6 ~(95.1,~90.2)~$~ \\ \hline

\end{tabular}
\end{table*}

\begin{figure}
\centering
\includegraphics[width=8cm]{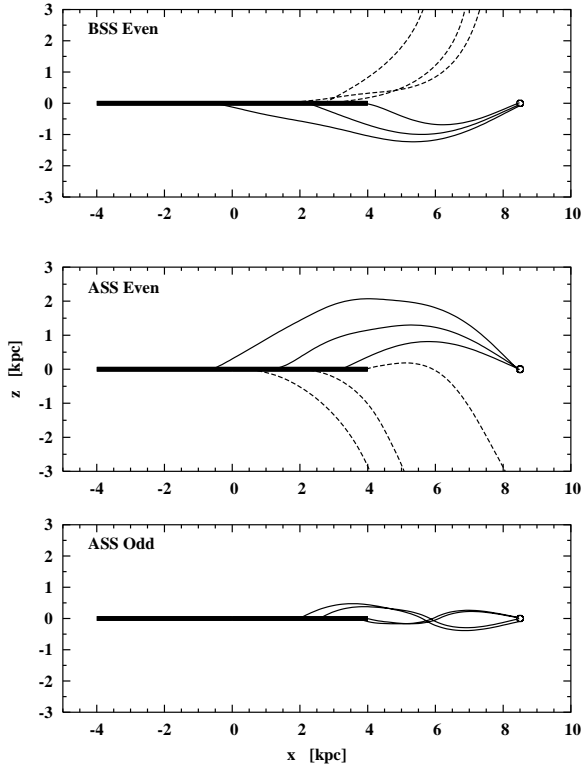}
\caption{
Examples of detected (solid lines) and non detected
(dashed lines) proton trajectories after travelling through different
GMF configurations. The trajectories are projected onto
a plane perpendicular to the Galactic disk that contains
the position of Solar system (plotted as a small open circle
at $x=8.5$ kpc) and the Galactic center (located at x=0, z=0). 
Proton energy is $E=10^{19}$ eV for all tracks. The thick solid line
parallel to the x-axis is the ring in the Galactic plane of radius $r=4$
kpc
around the Galactic center where the sources are located.
From top to bottom the GMF configuration corresponds to
BSS even parity, ASS even parity and ASS odd parity.
The random and dipole fields have been switched off to make this plot,
and we have used a 100 pc radius detector.
}
\label{fig:tracks}
\end{figure}

To understand how this strong anisotropy is actually realized, 
we show in Fig.\ref{fig:tracks} a sample of detected and non detected
proton trajectories at $10^{19}$ eV in the BSS and ASS (spiral field only)
even and odd parity configurations. What is shown is the projection 
of the proton trajectories through the GMF onto
a plane perpendicular to the Galactic disk containing the Solar 
system position and the Galactic center. In the BSS even parity 
configuration,
the GMF is directed towards $l\sim 270^\circ$ in the first arm 
that protons encounter on their paths to the 
Solar system as can be seen in Fig.\ref{fig:spiral-field}, 
so that their tracks tend to be concave. If a proton is 
injected from a source towards north, the GMF bends the trajectory so that it 
escapes from the Galaxy 
(dashed lines in the top panel of Fig. \ref{fig:tracks}).
If the proton is injected towards south, the GMF will bend its 
track towards north so that it may hit the Solar system and
will appear as coming from the Southern Galactic hemisphere
(solid lines in the top panel of Fig. \ref{fig:tracks}). The opposite behavior 
is true for the ASS even parity configuration, i.e. the tracks tend 
to be convex 
(middle panel in Fig.\ref{fig:tracks}) due to the GMF pointing towards Galactic
longitude $l\sim 90^\circ$ in the first magnetic arm between the 
sources and the Solar system (see right panel of
Fig.\ref{fig:spiral-field}). As a consequence only protons injected
towards north and bending back towards south can be detected, and will 
appear to come from the Northern Galactic hemisphere. 
If the field changes sign
across the Galactic plane (odd parity), both concave and convex trajectories
are possible and there is no prefered arrival direction. In fact typical
proton trajectories arriving at the detector cross the Galactic plane 
due to the change of sign of the GMF across the Galactic disk 
as can be seen in the bottom panel of Fig.\ref{fig:tracks}.

The use of a 1 kpc radius spherical detector, being the average distance to
the sources
in the Galactic ring of the order of $\sim 5$ kpc, induces an
unavoidable smearing in the proton arrival angles of $\sim 10^\circ$.
However this does not change 
our conclusions about the strong NS anisotropy we predict. 
The reason is that, for instance for the BSS even + $B_{\rm ran}$ + 
dipole configuration,
$~96\%$ of the events coming from south in the energy bin
$\log_{10}(E/{\rm eV})\in (18.3, 18.4)$ arrive from latitudes $b < -10^\circ$.
As a result the strong NS anisotropy given in Table\ref{tab:NS_BSS_even} 
is not expected to change significantly due to the 
$10^\circ$ smearing in arrival angle that may possibly shift 
their arrival directions so that they would actually appear to 
come from north.

\begin{acknowledgements}
We thank R.A. V\'azquez for helpful discussions.
This research is supported in part by NASA Grant ATP-0000-0080.
J. A-M is supported by the Spanish ``Ram\'on y Cajal" program,
and acknowledges the Xunta de Galicia (PGIDIT02 PXIC 20611PN),
and the MCYT (FPA 2001-3837, FPA 2002-01161 and FPA 2004-01198).
We thank the ``Centro de Supercomputaci\'on de Galicia" (CESGA)
for computer resources.
\end{acknowledgements}

\end{document}